# Dynamical patterns in individual trajectories toward extremism

Z. Cao[1,*], M. Zheng[1,*] Y. Vorobyeva[2], C. Song[1], N.F. Johnson[1]

[1]Complex Systems Initiative, Physics Department, University of Miami, FL 33146, U.S.A.

[2]Department of International Studies, University of Miami, Coral Gables, FL 33146, U.S.A.

* These authors contributed equally to this work

**Abstract.** Society faces a fundamental global problem of understanding which individuals are currently developing strong support for some extremist entity such as ISIS (Islamic State) – even if they never end up doing anything in the real world. The importance of online connectivity in developing intent has been confirmed by recent case-studies of already convicted terrorists. Here we identify dynamical patterns in the online trajectories that individuals take toward developing a high level of extremist support – specifically, for ISIS. Strong memory effects emerge among individuals whose transition is fastest, and hence may become 'out of the blue' threats in the real world. A generalization of diagrammatic expansion theory helps quantify these characteristics, including the impact of changes in geographical location, and can facilitate prediction of future risks. By quantifying the trajectories that individuals follow on their journey toward expressing high levels of pro-ISIS support -- irrespective of whether they then carry out a real-world attack or not – our findings can help move safety debates beyond reliance on static watch-list identifiers such as ethnic background or immigration status, and/or post-fact interviews with already-convicted individuals. Given the broad commonality of social media platforms, our results likely apply quite generally: for example, even on Telegram where (like Twitter) there is no built-in group feature as in our study, individuals tend to collectively build and pass through so-called super-group accounts.

**Significance Statement.** We identify dynamical patterns in the online trajectories that individuals take toward developing a high level of extremist support – specifically, for ISIS (Islamic State). Strong memory effects emerge among individuals whose transition is fastest, and hence may become 'out of the blue' threats. A generalization of diagrammatic expansion theory helps quantify these characteristics, including the impact of changes in geographical location, and can facilitate prediction of future risks. By focusing on the trajectories that individuals follow with respect to developing and expressing high levels of pro-ISIS support -- irrespective of whether they then carry out a real-world attack or not -- our findings can help move surveillance beyond reliance on static watch-list identifiers such as ethnic background or immigration status.

Relatively unsophisticated but high-impact attacks such as in Manchester, Stockholm, Paris and London in 2017, and Berlin, Nice, Brussels and Orlando in 2016, look set to become a fact of life *(1-3)* given the difficulty in detecting potential perpetrators who may act 'out of the blue' anywhere in the world. One extremist source is offering $50,000 to anyone, anywhere, just for attempting such an attack *(2)*. A fundamental problem faced by security agencies is how to move as far as possible 'left of boom' in order to detect individuals who are currently developing intent in the form of strong support for some extremist entity – even if they never end up doing anything in the real world. The importance of online connectivity in developing intent *(4-12)* has been confirmed by recent case-studies of already convicted terrorists by Gill and others *(6,7)*. Quantifying this online dynamical development can help move beyond static watch-list identifiers such as ethnic background or immigration status.

Here we address this problem using a unique dataset of online activity involving the global population of ~350 million users of the social media outlet VKontakte ([www.vk.com](www.vk.com)). VKontakte became the primary online social media source for ISIS propaganda and recruiting before moderator pressure forced activity toward encrypted alternatives such as Telegram in late 2015 *(13)*. Our experimental design and data gathering follows Ref. 12. Unlike Facebook which squashes such activity almost immediately, support for ISIS on VKontakte develops around online groups which are akin to everyday Facebook groups supporting a particular enterprise in sport, politics or education *(12)*. These online groups keep themselves open-source in order to attract new members, hence we are able to record current membership at every instant. Our hybrid system of application program interfaces (APIs) backed up by intensive manual cross-checking, shows that 91,781 of the ~350 million VKontakte users were members of at least one online pro-ISIS group (Fig. 1A) at the start of our study (1 Jan 2015) or became members during it (i.e. post 1 Jan 2015). Our method for determining a pro-ISIS group, together with explicit examples of its pro-ISIS content, are given in the Supplemental Information (SI). While our dataset is not provably complete or error-free, we believe that our approach of identifying online groups to unravel and quantify the trajectories of individual supporters *(12)* is as close as one can come without having to sift through all ~350 million users one by one, and without access to classified or private information. Given the broad commonality of social media platforms, the trajectory results that we report here likely apply more generally: for example, even on Telegram where (like Twitter) there is no built-in group feature, individuals tend to collectively build and pass through so-called super-group accounts *(11)*.

## Results

**Moderator bans:** Both individual accounts and groups become physically banned by VKontakte moderators when they become too extreme in their support of pro-ISIS violence. The act of banning produces a definite online announcement (e.g. Fig. 1B), hence it provides us with a well-defined measure of when an individual or group reaches a high level of pro-ISIS support. We can therefore unambiguously classify each individual in our dataset *while he/she is still developing* as a 'future-ban' individual (*b*) if he/she *in the future* reaches a high level of pro-ISIS support (i.e. become banned). Similarly, we can unambiguously classify each pro-ISIS group as a future-ban group (*B*) if it eventually gets banned (and *A* otherwise). During his/her development, each individual may join any number of *B* and/or *A* groups, and at any one moment may be fluctuating in terms of tending toward or away from this high level of extremist expression – and hence toward or away from becoming banned. Though our focus here is on the online development of these 'future-ban' individuals *irrespective* of whether they later carry out an extremist act or not, our analysis of media reports together with other individuals' mentions of usernames suggests that a significant number do. For example, the user list includes an eventual combatant who produced real-time audio recordings with street-level detail during assaults in Syria; an eventual suicide bomber who seems to have driven a truck of explosives into a Shia army in Iraq; an individual whose eventual combat activity in Iraq made headlines *(14)*; and an individual who transitioned to become the leader of Chechen fighters within ISIS.

**A** future-ban individual joins $n$'th group at time $t$ | future-ban individual joins $(n+1)$'th group at time $t' > t$

Example: user 1549171532 (male)
$B \to B \to B \to B \to B \to B \to B \to B \to B \to B \to B \to B \to B \to B \to B \to B \to B \to B \to B \to B \to A \to B \to B \to B \to A \to B \to B \to B \to B \to B \to B \to B \to B \to B \to B$

**B**

ВКонта́кте

has been suspended due to calls to violent actions

**C**

$$\Pi = P(0B) + P(1B) + P(2B) + ..$$

**D**

$P(nB)$ slope -2 $n$

**E**

$P(nB)$

$P(1B)$

$P(0B)$

$P(2B)$ $P(3B)$

$P(4B)$

$n$

**Figure 1: Individuals' online trajectories**. **A.** Schematic of group-joining in the complex pro-ISIS online space www.vk.com which comprises a mix of individuals (colored squares), groups (clouds), and content (examples show some less explicit postings). One observed sequence of joining events is listed as an example using abbreviation $B$ and $A$ for the joining of a 'future-ban' or 'no-future-ban' group respectively. **B.** Snapshot from VKontakte showing the measurable event of a group or individual getting banned. **C.** Exact diagrammatic expansion for the probability (i.e. risk) $\Pi$ which is the probability that any given individual in our study will eventually develop a sufficiently high level of pro-ISIS support that he/she violates VKontakte's rules against promoting pro-ISIS violence and hence his/her account gets banned. $P(nB)$ is the probability that his/her account will eventually get banned *and* he/she joins exactly $n$ future-ban ($B$) groups prior to this banning. Hence by definition, $P(nB) = 0$ for individuals whose accounts do not get banned. The probability propagator for a future-ban ($B$) group is shown as a gray strip. **D.** Log-log probability $P(nB)$, calculated from the empirical data by counting the fraction of individuals whose account eventually gets banned *and* who join exactly $n$ future-ban ($B$) groups between Jan 1 2015 and the instant his/her account gets banned. (By definition, $P(nB) = 0$ for no-future-ban individuals since they do not get banned). Line with slope $-2$ is shown as a guide. **E.** Same as **D** but over small $n$ range.

**Trajectories:** Of these 91,781 individuals, 7,707 later develop such extreme online support that their individual account becomes banned by VKontakte, i.e. there are 7,707 future-ban individuals. Hence the probability that a given developing individual will eventually reach a high level of extremist support (i.e. become banned) can be estimated from the frequency of occurrences in the data as $\Pi = \frac{7707}{91781} = 0.084$. Each developing individual's trajectory can be represented as a binary chain of group joinings (e.g. $B \rightarrow B \rightarrow A \rightarrow B \rightarrow A...$ as in Fig. 1A). The banning of groups occurs far more frequently than the banning of individuals *(12)*: It often happens that the collective content of a group quickly becomes very extreme, while the postings of any particular member may not. Hence the group gets banned at a given instant while the individual(s) does not. As a consequence, individuals who join lots of future-ban (*B*) groups are *not* necessarily the ones having their accounts banned. For example, the user in Fig. 1A joins 34 *B* groups and only 2 *A* groups, but never becomes banned because his/her individual postings – while actively supporting ISIS (see SI) -- are not judged to be sufficiently extreme. Indeed, among all the individuals who join 10 or more future-ban (*B*) groups, only 1,413 are future-ban individuals while 3,619 are not, meaning that the trajectory of future-ban individuals is not simply driven by the process of joining as many future-ban (*B*) groups as possible.

**Diagrammatic expansion:** Motivated by the physics approach to understanding transition probabilities in terms of successively higher-order interactions using diagrammatic expansion theory *(15)*, we unravel the contributions to the probability $\Pi$ that a given individual will develop a high level of extremist support (i.e. becomes banned and hence is a future-ban individual) by expanding $\Pi$ exactly in terms of successively higher-order interactions with *B* groups (Fig. 1C). The data (Fig. 1D) shows that the expansion terms $P(nB)$ for becoming banned after joining $n \geq 2$ future-ban groups are interrelated by an approximate power-law distribution $n^{-\alpha}$ with $\alpha \sim 2$ as opposed to a memoryless exponential decay. Though reminiscent of power-law distributions reported for everyday human activities, this is the first example related to extremist pathways. It is consistent with the idea that someone who has joined $n$ groups has accumulated $n$ potentially distinct narratives over time and hence needs to make $\sim n^2$ comparisons between pairs. This $\sim n^2$ increase suggests a similar increase in both required resources and potential narrative discord as $n$ increases, which in turn suggests that the probability $P(nB)$ may vary as $\sim n^{-2}$, as observed for $n \geq 2$. By contrast, Fig. 1E shows that $P(0B)$ -- and to a lesser extent $P(1B)$ -- is atypical. In particular, the empirical value of $P(0B)$ (i.e. the probability that an individual who gets banned will not join any future-ban groups between Jan 1 2105 and the banning of their account) is far smaller than that predicted by this simple power-law scaling. This has the important consequence that a risk estimate (i.e. probability) that a given individual will in the future reach a high level of pro-ISIS support (i.e. account becomes banned) based purely on observing 'lone wolf' individuals who subsequently join no $B$ groups (i.e. only using $P(0B)$) will be a significant underestimate since the $\sim n^{-2}$ scaling makes higher-order corrections to $P(0B)$ large in Fig. 1C. The fact that $P(1B)$ is by far the largest of all the expansion terms suggests that the impetus that a future-ban individual experiences toward becoming banned after joining his/her first future-ban group, is more influential than upon joining his/her second etc.

To explore these higher-order contributions to $\Pi$ more rigorously, we condition the probabilities on a particular value of the event-time lifetime $L_{\text{ban}}$, which is the number of pro-ISIS online groups of type $B$ and $A$ that a future-ban individual joins before becoming banned. Figure 2A shows $P(nB|L_{\text{ban}})$ for representative $L_{\text{ban}}$ values. (By definition, $P(nB|L_{\text{ban}})$ is undefined for individuals whose accounts do not get banned). The resulting empirical distributions deviate from the null model of a memoryless binomial process and provide evidence of a specific memory effect in the group-joining activity of those individuals who will eventually become banned – in particular, for small $L_{\text{ban}}$. This expansion unraveling of contributions to $\Pi$ – either conditioned to a given $L_{\text{ban}}$ or not – allows immediate prediction of all higher-order probability terms for an arbitrarily chosen individual, and hence the risk that this individual will reach a high level of pro-ISIS support, based simply on an approximate knowledge of the functional dependence of $P(nB)$. For example, evaluating $\Delta = (\Pi - P(0B))$ quantifies the impact online social media has on the probability (i.e. risk) that an individual will become extreme enough to have their account banned: using the numbers from our dataset, joining online groups enhances the probability that an individual eventually reaches a high enough level of pro-ISIS support that his/her account gets banned, from $P(0B) = \frac{1206}{91781} = 0.013$ to $\Pi = 0.084$ which is a huge increase of 546%. Evaluating the expansion for $\Pi$ analytically using additional knowledge of $P(1B)$ and applying the scaling factor from $P(1B)$ to $P(2B)$ for all $n$, reduces this error from 546% to 68%.

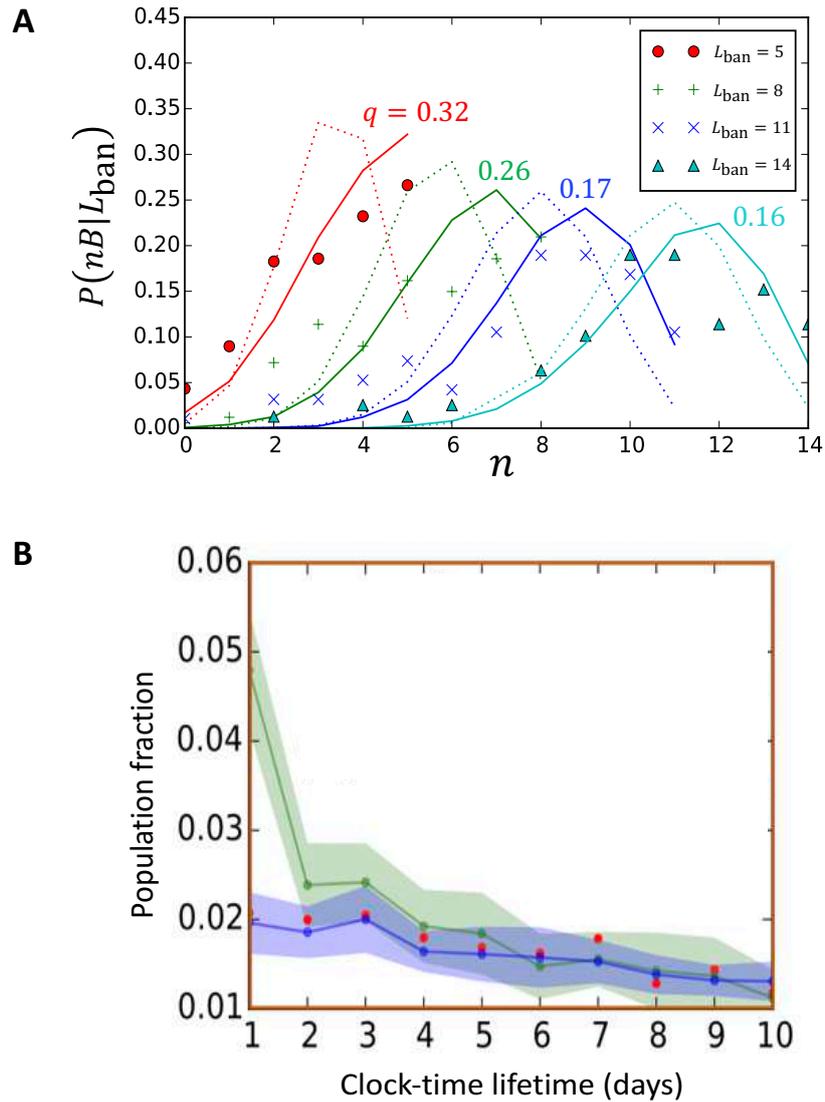

**Figure 2: Memory effects for shorter lifetimes. A.** Plot of conditional expansion terms $P(nB|L_{ban})$ given a specific event-time lifetime $L_{ban}$ for individuals who will eventually get banned (i.e. future-ban individuals). $P(nB|L_{ban}) = P(nB \cap L_{ban})/P(L_{ban})$ is the probability that any given individual in our study (i.e. out of all individuals) will get banned and that he/she joins exactly $n$ future-ban ($B$) groups prior to this banning, conditioned on his/her event-time lifetime until banning being $L_{ban}$. By definition, $L_{ban}$ and hence $P(nB|L_{ban})$ are undefined for individuals whose accounts do not get banned. Scattered points: empirical values obtained from the set of individual pathways in our dataset. Solid lines: our finite-memory model with parameter value $q$ shown and obtained from a maximum likelihood estimate (MLE). Dotted lines: result for a memoryless null model in which individuals have the same average group-joining rate as the data (obtained using MLE) hence yielding a binomial distribution. The largest memory effects arise for the shortest event-time lifetimes. **B.** Comparison of the clock-time lifetime $T_{ban}$ distribution from the empirical data (red dots), our mathematical model with memory effects (blue), and without memory effects (green). Error bands are obtained from simulations of the model. $T_{ban}$ is undefined for individuals whose accounts do not get banned. The largest memory effect arises for the shortest clock-time lifetimes.

**Modeling memory in lifetimes:** Figure 2A shows the results of a mathematical model that we introduce that captures these finite memory signatures in the empirical trajectories, and helps elucidate their nature. Each step in the model features a stochastic process in event-time in which the individual is assumed to join a group (with probability $q$) that is of the same type (i.e. $B$ or $A$) as the one that they joined in one of the past $m$ joining events. With probability $p(1-q)$ they join a $B$ group, and with probability $(1-p)(1-q)$ they join an $A$ group, where $p$ determines the individual preference of group types. Simulations show that the results are similar for all $m$ as long as $q < 0.7$, and that the theoretical $P(nB|L_{\text{ban}})$ is primarily determined by $q$. We therefore fix $m$ to be small and estimate $q$ and $p$ for each value of $L_{\text{ban}}$ from the empirical data using maximum likelihood estimates. The model agrees well with the data (Fig. 2A) and confirms that the largest impact of memory effects is for smallest $L_{\text{ban}}$. The fact that the memory effect decreases with an increase in event-time lifetime (Fig. 2A) is consistent with the notion that individuals that join many groups may have a less certain longer-term goal, and hence may act more randomly when choosing their next group type.

Any practical surveillance would also be mindful of clock-time. We now show that the existence of memory-effects in event-time lifetime does indeed carry over to clock-time. The data shows that event-time lifetimes ($L_{\text{ban}}$) act as a lower bound (see SI) for the corresponding clock-time lifetimes $T_{\text{ban}}$ – hence there are a spectrum of individual-dependent conversion factors between event-time and clock-time. This makes sense because individuals who visit a given number of online groups will likely differ considerably in how much clock-time they spend in and between visits: by contrast, event time just counts the cumulative number of groups joined. Despite this, Fig. 2B confirms that a strong memory effect indeed arises for individuals with short clock-times (now $T_{\text{ban}}$) and that these memory effects can still be modeled using a continuous-time version of our memory-dependent model from Fig. 2. During any given day, an individual may have no group-joining event, hence the stochastic process mimics a walk in an abstract ideological space as opposed to a group-joining space. We hence consider the simplest case of a memoryless one-dimensional walk in which an individual gradually moves toward, or away from, an absorbing boundary that defines their exit from the system (i.e. account gets banned) and hence determines their clock-time lifetime $T_{\text{ban}}$. As $T_{\text{ban}}$ decreases, the empirical data shows an increasing deviation from this memoryless walk result. However, when we add a similar type of memory as before, i.e. with probability $q$ the individual takes an action (step) that copies the previous change, we find that the model with memory does now fit the data well (Fig. 2B).

An immediate practical implication of our findings so far is that strong, observable memory effects occur for the shortest lifetime individuals, whether measured in event-time or clock-time. This is fortunate since individuals who are moving quickly toward becoming banned (and hence toward showing a high level of pro-ISIS support) are also likely to be of most interest as potential threats. Though they may end up never carrying out a real-world extremist act, their banned status means that they likely harbor the strongest intent.

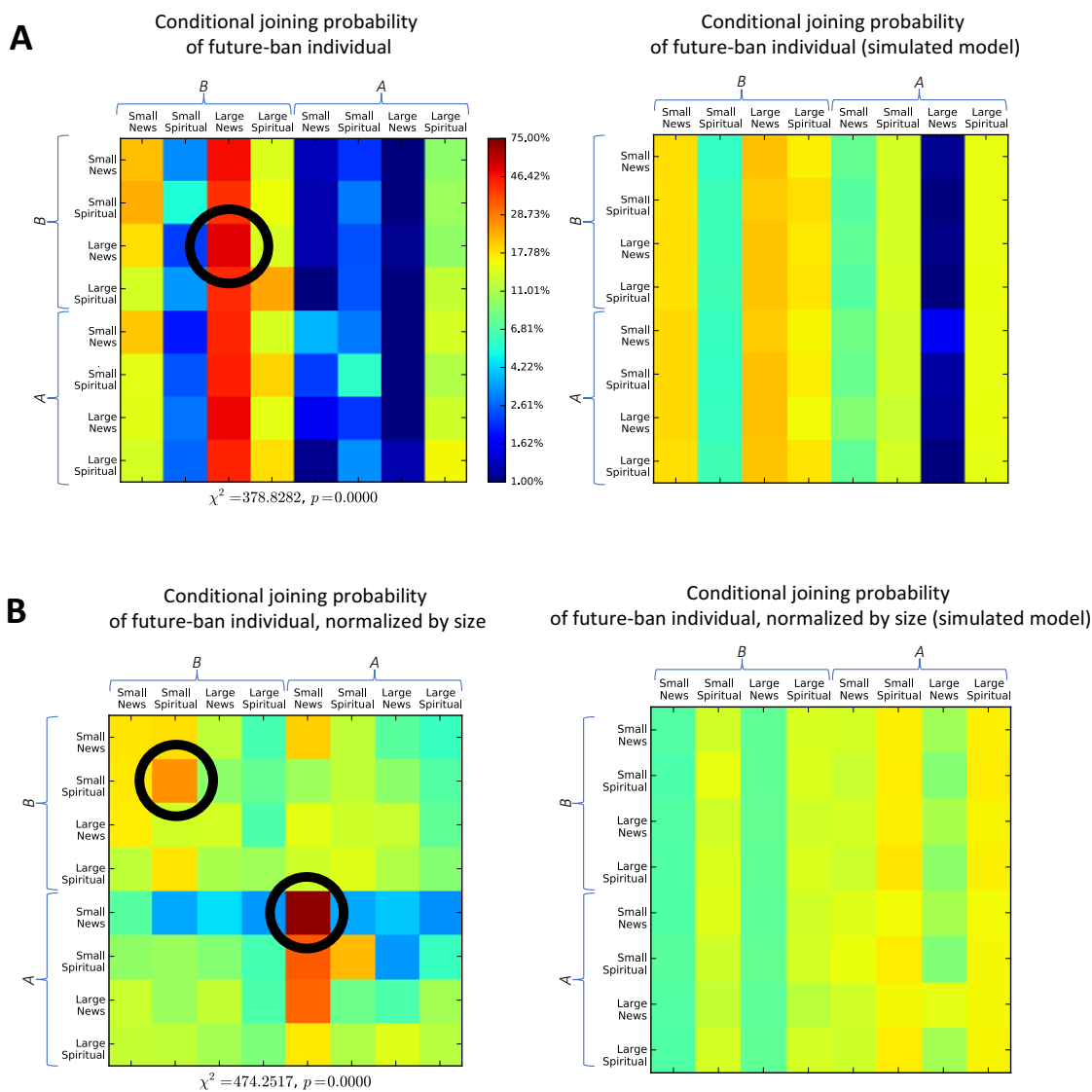

**Figure 3: Group joining**. **A**. Transition matrix obtained from the empirical data (left) and a simulation (right, see SI) showing conditional probability that a future-ban individual in a group of class $i$ (row) will next join a group of class $j$ (column). Future-ban individuals have the greatest preference for next joining a banned, large, news group, with this type of group acting as a global attractor (black ring). **B**. Corresponding result after renormalizing by the number and size of groups in class $j$. Two additional weaker attractors now emerge (black ring).

**Preferred dynamical transitions:** In accordance with the existence of memory effects in the trajectory of group joinings, we find that future-ban individuals' choice of their next group to join depends on the last group that they joined, and that there are rather well-defined attractors. The classes of groups are designated according to their size compared to the median group size ('small' or 'large') and whether they have a majority of postings that are focused on a news story ('news') or on more abstract discussions (we call these 'spiritual', since most of these are indeed spiritual in their content). Though other classifiers are in principle possible, and this choice of classifiers is somewhat subjective, we have checked that only using a subset does not change our main conclusions, and instead muddles the behavioral patterns. Figure 3A (left) uses the relative frequencies of online group-joining events to generate estimates of the conditional probability that a future-ban individual who most recently chose a group of class $i$ (row) will next join a group of class $j$. The SI provides details of the calculation and additional results, including for subsets of attributes. Figure 3A (right) shows the corresponding result if these individuals were to choose their next group based purely on the number and size of groups that exist at the time of their choice – i.e. akin to preferential attachment as in the coalescence-fragmentation model in Ref. 12.

Figure 3A shows that future-ban individuals have the greatest preference for next joining a banned, large, news group, with this type of group acting as a global attractor. Although this in part reflects the higher relative number and size of this group class, the differences between Fig. 3A left and right panels (together with the very high $\chi^2$ and negligible $p$-values shown) show that the full story behind these individuals' choice-making lies beyond a pure size effect. To explore this further, we renormalize the transition probabilities by the number of groups and members in the future class $j$, yielding the results in Fig. 3B. The simulated model for individuals who are eventually banned (Fig. 3B right) still shows small fluctuations because of the finite time window, but the empirical result (Fig. 3B left) shows marked differences that are statistically significant ($\chi^2 = 474.3$, $p < 0.00005$). Figure 3B (left) shows that future-ban individuals have an additional attraction, beyond preferential attachment, toward either future-ban, small, spiritual groups or no-future-ban, small, news groups. A theoretical model that incorporates 'character' into preferential attachment – as presented in Ref. 22 – could help tease out these differences, backed up by more detailed knowledge of individuals' circumstances.

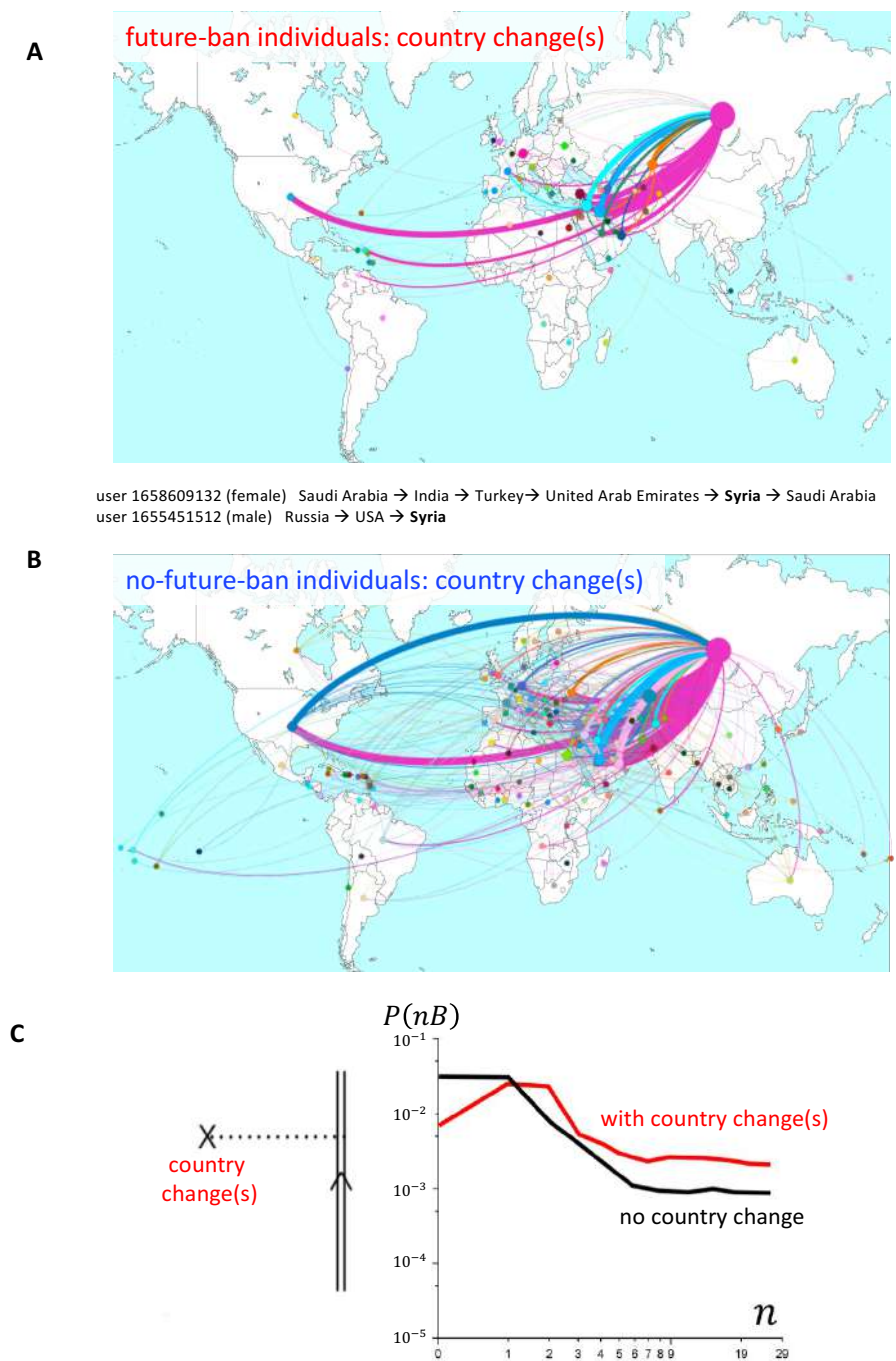

**A** future-ban individuals: country change(s)

user 1658609132 (female)  Saudi Arabia → India → Turkey→ United Arab Emirates → **Syria** → Saudi Arabia
user 1655451512 (male)  Russia → USA → **Syria**

**B** no-future-ban individuals: country change(s)

**C** country change(s)

$P(nB)$

with country change(s)

no country change

$n$

**Figure 4: Influence of geography. A.** Flow of future-ban individuals based on reported geographical locations in online profiles. The flow from country X to country Y is the number of individuals that change location from X to Y, and is proportional to the thickness of the line. The line colour corresponds to the originating country X. Each country's node is at its geographic centre. Two individuals' trajectories are listed as examples. Although Russia resembles a travel hub for all individual types because VKontakte has a large number of Russian users, many individuals' pathways do not pass through it (e.g. user 1658609132). **B.** Same as **A** but for individuals whose accounts are not eventually banned ('no-future-ban' individuals). For future-ban individuals (**A**), this country return rate $r = 21.6\%$ while for no-future-ban individuals (**B**) $r = 44.7\%$, which is more than twice. **C.** Scattering between countries is associated with a shift in the dependence of $P(nB)$. Red line is for future-ban individuals who changed country at least once. Black line is for future-ban individuals who never changed country.

**Extension to location:** The diagrammatic expansion (Fig. 1C) also opens up a new pathway toward addressing the important practical question of how an individual's online pathway interacts with their real-world movements (Fig. 4A-B). Geographical location is a declared quantity in an individual's online profile. A change in an individual's declared geographical location may act as a 'scattering' event in terms of changing the online trajectory of that individual (Fig. 4C) because, for example, of a change in his/her attitude or circumstances. In a physical system *(15)*, such scattering is known to enhance or diminish the likelihood of particular future trajectories. Of course, more data is required to rigorously corroborate any individual's specific country movements; and hence to link these to specific changes in the diagrammatic expansion in Fig. 4C; and ultimately obtain more accurate predictions of the risk of developing a high support level (i.e. account gets banned). However, during a manual inspection of postings that include backgrounds from photos, we did not find any that were overtly inconsistent with a reported country's expected environment. At the very least, the location that an individual declares can presumably be regarded as reflecting the location with which that individual currently identifies his/her activity.

Taking our current data at face value, and with these caveats in mind, the VKontakte data shows (Figs. 4A-B) that there are visible patterns in how individuals of the two different types change their declared locations, with future-ban individuals visibly avoiding loops (i.e. less return trips). We can quantify this in a simple way by looking at the fraction of return trips. If $m$ future-ban individuals change country from $i$ to $j$ during our period of study, and $n$ individuals change country from $j$ to $i$, we can define a country return rate for this country pair as $\min(m,n)/\max(m,n)$. We then average this quantity over all possible pairs of countries to obtain a country return rate $r$. For future-ban individuals (Fig. 4A), this country return rate $r = 21.6\%$ while for no-future-ban individuals (Fig. 4B) $r = 44.7\%$, which is more than twice. Figure 4C shows that for future-ban individuals that change country, their higher-order ($n \geq 2$) probabilities tend to *increase* in the diagrammatic expansion, i.e. estimates of the probability of interest $\Pi$ based solely on $P(0B)$ will be worse. This suggests an opportunity to extend existing research on real-space conflict and mobility *(16-22)* to develop a fuller theory of extremist dynamics in coupled cyber-physical space. Among future-ban individuals, the average number of online pro-ISIS groups joined immediately *after* a move from Syria to any other country, is 0.36 as compared to an average over time of 0.18. This is a statistically significant increase ($p = 0.03$). Similarly, the average number of online groups that such future-ban individuals join immediately *before* a move to Germany from any other country is 0.48, as compared to an average over time of 0.36 ($p = 0.08$). The case of U.S. is similar to Germany ($p = 0.1$). This means that among future-ban individuals who come from anywhere in the world, the ones that enter Germany and U.S. tend to do so immediately after a burst of online group-joining activity – a fact that could be used to identify future high-risk individuals in those countries. Among all those who are at some stage in Syria, there is a higher percentage of future-ban individuals who go from Syria to France than go to Syria from France. The same is true for Turkey, Iraq and Australia. The vast majority of individuals moving from Syria to U.S. or U.K. are no-future-ban, suggesting that the threat in U.S. and U.K. is a long-term latent one – perhaps like the 2017 Manchester bomber. With more reliable country-specific data, future risk probabilities tailored to these specific locations and movements could be calculated by conditioning the scattering transitions in Fig. 4C.

## Conclusions

Our paper addresses the pressing societal problem of understanding which individuals are currently developing intent in the form of strong support for some extremist entity – even if they never end up doing anything in the real world. Using a unique dataset from an online social media source, we have identified specific dynamical patterns in the online trajectories that individuals take toward developing a high level of extremist support – specifically, for ISIS (Islamic State). We identified strong memory effects emerging among individuals whose transition is fastest, and hence may become 'out of the blue' threats in the real world. A generalization of diagrammatic expansion theory helped quantify these characteristics, including the impact of changes in geographical location, and can now be used to facilitate prediction of future risks. Our findings help move beyond post-fact interviews with already-convicted terrorists, by focusing on the trajectories that individuals follow with respect to developing and expressing high levels of pro-ISIS support -- irrespective of whether they then carry out a real-world attack or not. Our findings also help move global security debates beyond static watch-list identifiers such as ethnic background or immigration status. Given the broad commonality of social media platforms, the results that we report here likely apply more generally: for example, even on Telegram where (like Twitter) there is no built-in group feature, individuals tend to collectively build and pass through so-called super-group accounts *(11)*.

**References**

1. M. Birnbaum, The Washington Post, Europe may face a grim future with terrorism as a fact of life. December 24, 2016. Available at https://www.washingtonpost.com/world/europe/after-berlin-market-a...536-c84c-11e6-acda-59924caa2450_story.html?utm_term=.e03c1bbad77c

2. IS group to step up attacks on Europe – Europol. BBC News http://www.bbc.com/news/world-europe-38179653. Accessed 2 Dec, 2016

3. D. Hambling. How Islamic State is using consumer drones. December 9, 2016. Available at http://www.bbc.com/future/story/20161208-how-is-is-using-consumer-drones

4. BBC News. Russian students targeted as recruits by Islamic State. http://linkis.com/dmFgu (July 24, 2015)

5. P. Gill, *Lone Actor Terrorists: A Behavioural Analysis*. (Routledge, London, 2015)

6. P. Gill, J. Horgan, P. Deckert. Bombing alone: Tracing the motivations and antecedent behaviors of lone-actor terrorists. *Journal of Forensic Sciences* **59**, 425– 435 (2014)

7. P. Gill, E. Corner. Lone-actor terrorist use of the Internet and behavioural correlates. In *Terrorism Online: Politics, Law, Technology and Unconventional Violence*. Eds. L. Jarvis, S. Macdonald & T.M. Chen (Routledge, London, 2015)

8. L. Von Behr, A. Reding, C. Edwards, L. Gribbon. *Radicalization in the Digital Era: The Use of the Internet in 15 Cases of Terrorism and Extremism*. (RAND, Santa Monica CA, 2013). Available at rand.org/content/dam/rand/pubs/research_reports/RR400/RR453/RAND_RR453.pdf

9. P. Gill, E. Corner, M. Conway, A. Thornton, M. Bloom, J. Horgan. Terrorist Use of the Internet by the Numbers Quantifying Behaviors, Patterns, and Processes. *American Society of Criminology* (2016, in press)

10. J.N. Shapiro. *The terrorist's dilemma: Managing violent covert organizations* (Princeton University Press, Princeton, 2013)

11. J.M. Berger, H. Perez. The Islamic States diminishing returns on Twitter. GW Program on Extremism 2 (2016). Available at https://cchs.gwu.edu/sites/cchs.gwu.edu/files/downloads/Berger_Occasional%20Paper.pdf

12. N.F. Johnson, M. Zheng, Y. Vorobyeva, A. Gabriel, H. Qi, N. Velasquez, P. Manrique, D. Johnson, E. Restrepo, C. Song, S. Wuchty, S. New online ecology of adversarial aggregates: ISIS and beyond. *Science* **352**, 1459-1463 (2016)

13. J. Paraszczuk. Why Are Russian, Central Asian Militants Vanishing From Social Networks? RadioFreeEurope November 05, 2015. Available at http://www.rferl.org/a/russian-central-asian-militants-vanish-social-networks/27347535.html

14. http://www.rferl.org/a/why-the-death-of-a-chechen-media-activist-in-iraq-suggests-is-may-be-in-trouble-/26974272.html Accessed 12-9-2016

15. R.D. Mattuck. A Guide to Feynman Diagrams in the Many-Body Problem (Dover, New York, 1992)

16. M.C. González, C.A. Hidalgo, A.L. Barabási. Understanding individual human mobility patterns. *Nature* **453**, 779-782 (2008) Accessed 12-9-2016

17. G.M. Viswanathan, G.E. da Luz, E.P. Raposo, H.E. Stanley. *The Physics of Foraging*. (Cambridge University Press, Cambridge, 2011)

18. I.D. Couzin, J. Krause, N.R. Franks, S.A. Levin. Effective leadership and decision-making in animal groups on the move. *Nature* **433**, 513-516 (2005) doi:10.1038/nature03236

19. S. Gavrilets. Collective action and the collaborative brain. *J. R. Soc. Interf*ace **12**, 20141067 (2015) http://dx.doi.org/10.1098/rsif.2014.1067

20. R. Wrangham, L. Glowacki. Intergroup aggression in chimpanzees and war in nomadic hunter-gatherers. *Human Nature* **23**, 5 (2012)

21. M.L. Wilson et al. Lethal aggression in Pan is better explained by adaptive strategies than human impacts. *Nature* **513**, 414–417 (2014)

22. L. Glowacki, A. Isakovc, R.W. Wrangham, R. McDermott, J.H. Fowler, N.A. Christakis. Formation of raiding parties for intergroup violence is mediated by social network structure. *Proc. Natl. Acad. Sci.* **113**, 12114-12119 (2016). Available at www.pnas.org/cgi/doi/10.1073/pnas.1610961113

**Acknowledgements** We are grateful to Andrew Gabriel and Anastasia Kuz for initial help with data collection and analysis, and to Jim Nearing and Thom Curtright for discussions. NFJ gratefully acknowledges funding under National Science Foundation (NSF) grant CNS1522693 and Air Force (AFOSR) grant FA9550-16-1-0247. The views and conclusions contained herein are solely those of the authors and do not represent official policies or endorsements by any of the entities named in this paper. Data described in the paper are presented in the file *XXX.xlsx* that is available in the SI.

**Supplemental Information (SI)**